\documentclass[
reprint,
superscriptaddress,
amsmath,
amssymb,
aps,
pra,
floatfix,
]{revtex4-2}

\usepackage{graphicx}
\usepackage{dcolumn}
\usepackage{bm}
\usepackage{soul}
\usepackage{braket}
\usepackage[colorlinks=true,urlcolor=blue,citecolor=blue,linkcolor=blue]{hyperref}
\usepackage[dvipsnames]{xcolor}
\usepackage{mathtools}
\usepackage{array}
\usepackage{enumerate}   
\usepackage{siunitx}
\usepackage{lipsum}
\usepackage[normalem]{ulem}

\DeclareMathOperator{\pump}{\mathrm{p}}
\DeclareMathOperator{\signal}{\mathrm{s}}
\DeclareMathOperator{\idler}{\mathrm{i}}

\begin{document}

\preprint{APS/123-QED}


\title{High-Dimensional Maximally Entangled OAM States by Joint Pump and Modal-Space Engineering}





\author{Richard Bernecker}
 \email{richard.bernecker@uni-jena.de}
\affiliation{%
 Theoretisch-Physikalisches Institut, Friedrich-Schiller-Universität Jena, Max-Wien-Platz 1, D-07743 Jena, Germany}
\affiliation{
 Helmholtz-Institut Jena, Fröbelstieg 3, D-07743 Jena, Germany 
}
 
\author{Baghdasar Baghdasaryan}
\affiliation{%
Institut für Angewandte Physik, Friedrich-Schiller-Universität Jena, Albert-Einstein-Str. 6, D-07745 Jena, Germany
}

\author{Fabian Steinlechner}
\affiliation{%
Institut für Angewandte Physik, Friedrich-Schiller-Universität Jena, Albert-Einstein-Str. 6, D-07745 Jena, Germany
}
\affiliation{
Fraunhofer-Institut für Angewandte Optik und Feinmechanik Jena, Albert-Einstein-Str. 7, D-07745 Jena, Germany
}

\author{Stephan Fritzsche}
\affiliation{%
 Theoretisch-Physikalisches Institut, Friedrich-Schiller-Universität Jena, Max-Wien-Platz 1, D-07743 Jena, Germany}
\affiliation{
 Helmholtz-Institut Jena, Fröbelstieg 3, D-07743 Jena, Germany 
}
\affiliation{
GSI Helmholtzzentrum für Schwerionenforschung GmbH, Planckstrasse 1, D-64291 Darmstadt, Germany
}

\date{\today}

\begin{abstract}
Spontaneous parametric down-conversion (SPDC) generates correlated photon pairs that inherently exhibit high-dimensional spatial entanglement in orbital angular momentum (OAM) modes. In practice, the resulting two-photon states are typically not maximally entangled due to a highly non-uniform distribution of generated OAM modes. Although pump-beam engineering has emerged as a tool to prepare high-dimensional entangled states, it has usually been restricted to low-dimensional subspaces, i.e., small sets of selected OAM modes for each photon within the full infinite-dimensional two-photon Hilbert space. A general understanding of the requirements for the generation of maximally entangled states (MESs) for any dimensionality $d$ with $d>2$ is lacking. Here, we demonstrate how pump engineering enables the generation of high-dimensional MESs for arbitrary $d$ defined in $d \times d$ subspaces. By identifying the interplay between the spatial pump structure and the choice of OAM subspace, the required pump profile for a target MES is directly determined. Our results reveal that maximal entanglement with OAM requires a careful and nontrivial design of the effective high-dimensional Hilbert subspace. 
\end{abstract}
\maketitle

\newpage

\twocolumngrid

\begin{figure*}
    \centering
    \includegraphics[width=0.95\linewidth]{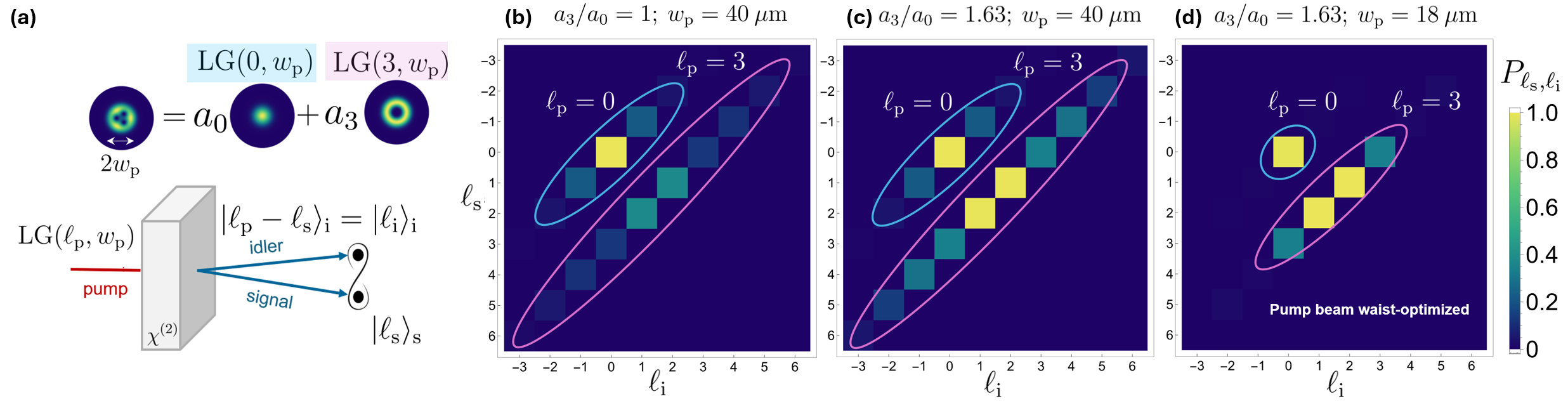}
    \caption{Sketch of the SPDC process and the pump-imprinted OAM spectra. (a) During SPDC, the spatial structure of a pump field formed by a coherent superposition of LG modes with different OAM indices $\ell_{\pump}$ is transferred to the OAM correlations of the generated photon pairs. The two-photon state of signal and idler is a coherent, non-separable superposition of OAM modes $\ket{\ell}_{\signal}$ and $\ket{\ell_{\pump}-\ell}_{\idler}$, constrained by OAM conservation. (b) The corresponding normalized OAM spectrum of a superposition of pump modes $\ell_{\pump}=0$ (blue) and $\ell_{\pump}=3$ (pink) with the characteristic anti-diagonal correlations in the OAM spectrum. For an equally weighted superposition, the maximum amplitude along the blue diagonal does not match the maximum amplitude along the pink diagonal. (c) If the relative weight of $a_3$ compared to $a_0$ is increased, the maximum amplitudes along the two anti-diagonals can be equalized. (d) An appropriate choice of pump waist $w_{\pump}$ minimizes the spiral bandwidth, concentrating the population around the two-photon modes $\ket{0,0}$ (blue) or $\ket{1,2}$ and $\ket{2,1}$ (pink).} 
    \label{figure01}
\end{figure*}

\section{Introduction}
Photons are a versatile platform for quantum information \cite{slussarenko2019photonic, flamini2019photonic, zhong2020quantum}. High-dimensional entangled states (HDES), where each photon is described by a Hilbert space of dimension $d>2$ \cite{erhard2020advances}, attract interest for increased information capacity and enhanced robustness in quantum communication \cite{cozzolino2019high, zhu2021high}, reduced circuit complexity in quantum computing \cite{wang2020qudits}, and increased phase sensitivity in quantum metrology \cite{dambrosio2013photonic, jha2011supersensitive}.


Transverse spatial modes of light, particularly orbital angular momentum (OAM) modes, provide a discrete and unbounded basis for high-dimensional photonic entanglement \cite{allen1992orbital, molina2007twisted, yao2011orbital, erhard2018twisted, forbes2019quantum}. OAM-based states are used in high-capacity free-space communication \cite{krenn2014communication, mirhosseini2015high, krenn2015twisted}, dense multiplexing \cite{wang2012terabit, bozinovic2013terabit}, and superresolution imaging \cite{moreau2019imaging}. Photon pairs generated via spontaneous parametric down-conversion (SPDC) \cite{anwar2021entangled} intrinsically exhibit high-dimensional entanglement in the spatial degree of freedom based on the conservation of OAM \cite{mair2001entanglement, fickler2012quantum}. 

The generated OAM modes are typically non-uniformly distributed, yielding non-maximally entangled two-photon states \cite{di2010measurement, krenn2014generation}. However, many quantum-information protocols are optimized for, or explicitly formulated in terms of, maximally entangled states (MESs), the high-dimensional analogues of Bell states \cite{horodecki2009quantum}. MESs maximize the entanglement entropy, uniformly exploit the full available Hilbert space, and serve as ideal resources to achieve optimal fidelity and error resilience in quantum-information protocols.

Crystal engineering \cite{xu2021manipulating, rozenberg2022inverse, bernecker24high} can be applied to shape the OAM spectrum. However, this technique usually requires the fabrication of tailored nonlinear media that are optimized for fixed dimensionality $d$. Furthermore, post-generation schemes like entanglement concentration \cite{vaziri2003concentration, dada2011experimental, wang2017generation, chen2020coherent} can flatten the measured OAM spectrum, but fail to achieve genuine at-source state engineering and add complexity to the setup. 

Furthermore, pump-beam engineering, e.g., via spatial light modulators (SLMs), enables programmable control of the OAM spectrum and has proven to be an experimentally convenient tool for preparing HDES \cite{liu2020increasing, boucher2021engineering, torres2003preparation}. Yet, the generation of arbitrary-dimensional MESs remains elusive: prior works are often constrained to small-dimensional subspaces \cite{yao2011angular, kovlakov2018quantum, liu2018coherent} or rely on technically demanding optimization procedures of the pump profile \cite{bornman2021optimal,xu2022manipulating, karan2023postselection, comer2026inverse}, without establishing general conditions for maximal entanglement.

In this paper, we introduce a broadly applicable approach for generating MESs in $d\times d$-dimensional OAM subspaces. Our method jointly engineers the pump beam and local OAM subspaces to establish conditions for maximal entanglement without numerical optimization of the pump profile. Our analytical construction is not restricted to a specific phase-matching type or nonlinear medium. Moreover, the proposed scheme ensures the spectral and spatial purity of the generated state.
Thus, our pump-engineering approach provides, together with modal postselection, a systematic construction to generate pure MESs in flexibly chosen OAM subspaces of, in principle, arbitrary dimensionality. 

\section{Theoretical Background}

\subsection{Two-photon entanglement in OAM}

Since the pioneering experiment by Mair \textit{et al.} \cite{mair2001entanglement}, which demonstrated OAM conservation and entanglement in the SPDC process, many works have studied the controlled generation of high-dimensional entanglement, recently even at the single-photon level \cite{kopf2025conservation}. To analyze OAM entanglement in SPDC, the biphoton state is conveniently expanded in an eigenbasis $\ket{\ell}$ of the OAM operator $\hat{L}$, i.e., $\hat{L}\ket{\ell} = \ell \hbar \ket{\ell}$ with $\ell \in \mathbb{Z}$. Laguerre–Gaussian (LG) modes provide a natural basis for this decomposition \cite{walborn2010spatial}. For generality, we assume that the pump beam is also described by an LG mode. Consequently, the entangled state of a signal ($\signal$) and idler ($\idler$) photon from SPDC
driven by an LG pump beam ($\pump$) carrying an OAM $\ell_{\pump}$ in the paraxial, quasi-collinear regime (cf. Fig.~\ref{figure01}(a)) is
\begin{align}
    \ket{\psi} &= \sum_{\ell_{\signal}} C_{\ell_{\signal}}(\ell_{\pump}, w_{\pump})  \; \underbrace{
    \ket{\ell_{\signal}} \ket{\ell_{\pump}-\ell_{\signal}}}_{:=\ket{\ell_{\signal},\ell_{\idler}}},
    \label{StateLGpump}
\end{align}
where OAM conservation sets $\ell_{\idler}=\ell_{\pump}-\ell_{\signal}$. The expansion amplitudes $C_{\ell_{\signal}}$ depend on the experimental geometry, the nonlinear crystal, and the pump beam \cite{miatto2011full, karan2020phase, bernecker2023spatial, torres2004quasi, dosseva2016shaping}. Since we aim to shape the pump beam parameters to design HDES, we explicitly emphasize its waist $w_{\pump}$ and OAM $\ell_{\pump}$. The OAM conservation directly correlates the signal and idler modes. The projection probability for the two-photon mode 
$\ket{\ell_{\signal},\ell_{\idler}}$ 
is given by $P_{\ell_{\signal},\ell_{\idler}}(\ell_{\pump}, w_{\pump}) := |C_{\ell_{\signal}}(\ell_{\pump}, w_{\pump})|^2$. Here, we consider degenerate SPDC where the wave numbers satisfy $k_{\pump}/2 = k_{\signal} = k_{\idler}$. This, combined with equal projection waists for signal and idler, enforces the symmetry condition $C_{\ell_{\signal}} = C_{\ell_{\pump}-\ell_{\signal}}$ \cite{bernecker24high}. In addition to the azimuthal index $\ell$, Laguerre–Gaussian modes are characterized by a radial index $p$ \cite{plick2015physical}. Different OAM components can be correlated with distinct radial-mode states. Tracing over unresolved radial degrees of freedom can therefore reduce the purity of the OAM state. Our main results are formulated for signal and idler modes selected at zeroth radial order,
$p_s=p_i=0$. The detailed description of the state in the broadband regime can be found in Appendix \ref{Appenndix: Details on the biphoton state} and higher-order radial correlations \cite{plick2015physical, fontaine2019laguerre, gu2018gouy, zhou2017sorting, fu2018realization} in Appendix \ref{Appenndix:Expansion amplitude of the full LG basis}. The same analytical framework and engineering
procedure also apply in the broadband regime, where the
pump weights are determined from frequency-integrated
coincidence probabilities. Moreover, the target mode pairs
in our construction share the same 
spectral dependence under the assumptions of our model. The conditional target state therefore retains its spatial purity after tracing
over frequency \cite{baghdasaryan2022generalized} as described in Appendix \ref{Appendix: Relative pump weights}.

\subsection{Pump-beam engineering}

Central to our approach is the freedom to utilize a general spiral distribution for the pump. Mathematically, this is expressed as an arbitrary superposition of LG modes \cite{bernecker24high}
\begin{align*}
V_{\pump} = \sum_{\ell_{\pump}}   a_{\ell_{\pump}} \: \text{LG}_{}^{}(\ell{_{\pump}}, w_{\pump}).
\end{align*}
with $\sum_{\ell_{\pump}} | a_{\ell_{\pump}}|^2=1$. Such a pump profile can be readily realized using commercially available SLMs \cite{zhu2014arbitrary}. In the low-gain regime of SPDC, the generated two-photon amplitude is linear in the pump field \cite{baghdasaryan2022generalized}. Therefore, the two-photon state generated by a pump superposition is the coherent sum of the contributions associated with each pump OAM component,
\begin{equation}
    \ket{\tilde{\psi}} = \sum_{\ell_{\text{p}}} a_{\ell_{\text{p}}} \; \sum_{\ell_{\signal}} C^{}_{\ell_{\signal}}(\ell_{\pump}, w_{\pump})  \; \ket{\ell_{\signal} , \ell_{\idler}}
    . 
    \label{Fullstate}
\end{equation}


Each pump component with $\ell_{\pump}$ populates one \textit{anti-diagonal} in the OAM spectrum, the distribution of $P_{\ell_{\signal},\ell_{\idler}}$,  while the coefficients $a_{\ell_{\pump}}$ control the relative weights between different anti-diagonals. Fig. \ref{figure01} (b) shows the normalized OAM spectrum for an equally weighted superposition of the pump modes $\mathrm{LG}(0,w_{\pump})$ and $\mathrm{LG}(3,w_{\pump})$. The resulting spectrum consists of two anti-diagonals, shown in blue and pink, each satisfying OAM conservation for the corresponding pump component. In contrast to the single-pump scenario, the signal and idler OAM modes are no longer connected by a single anti-correlation rule: a given signal mode can be paired with different idler modes belonging to different pump anti-diagonals.

Equal pump weights, $a_0=a_3$, do not generally produce equal two-photon amplitudes on the two anti-diagonals. The pump weights $a_{\ell_{\pump}}$ can therefore be adjusted to rescale the contributions of the different anti-diagonals. To equalize different OAM amplitudes, the calculation of the pump weights can be easily obtained via a matrix equation for the state coefficients (see Appendix \ref{Appendix: Relative pump weights}). For example, the coefficient $a_0 = (\sqrt{d} \; C_0(0,w_{\pump}))^{-1}$ enforces an amplitude $1/\sqrt{d}$ for the state $\ket{0,0}$. In Fig. \ref{figure01} (c), the weights are chosen such that the maxima of each anti-diagonal are equalized. 



Each anti-diagonal represents an independent distribution of possible modes, characterized by a width, called spiral bandwidth \cite{torres2003quantum}. The spiral bandwidth can be reduced by tuning the pump waist, concentrating the distribution near its dominant amplitudes \cite{kovlakov2017spatial}, as shown in Fig.~\ref{figure01} (d). The dominant amplitudes occur at the center of this anti-diagonal. Usually, this setting is referred to as the Single-Schmidt mode regime (SSMR). Here, higher-radial orders are minimized too \cite{kovlakov2017spatial, kovlakov2018quantum}. For even $\ell_{\pump}$, the strongest contribution is the symmetric mode pair with OAM $\ket{\frac{\ell_{p}}{2},\frac{\ell_{p}}{2}}$.
For odd $\ell_{p}$, the center lies between two OAM modes, so the two neighboring central amplitudes of the modes $\ket{\frac{\ell_{p}+1}{2},\frac{\ell_{p}-1}{2}}$ and $\ket{\frac{\ell_{p}-1}{2},\frac{\ell_{p}+1}{2}}$
are degenerate and maximal as shown in Fig.~\ref{figure01} (b) and (c). For $|\ell_{\pump}|\geq 2$, the finite spiral bandwidth gives rise to additional neighboring modes with non-negligible amplitudes, which cannot be completely suppressed by pump shaping alone \cite{bernecker2025engineering}. Thus, while $\ell_{\pump}=0$ yields a strongly populated mode $\ket{0,0}$, higher pump modes produce broader spectra with non-vanishing contributions, as evident from Fig. \ref{figure01} (c).

\begin{figure*}
    \centering
    \includegraphics[width=0.92\linewidth]{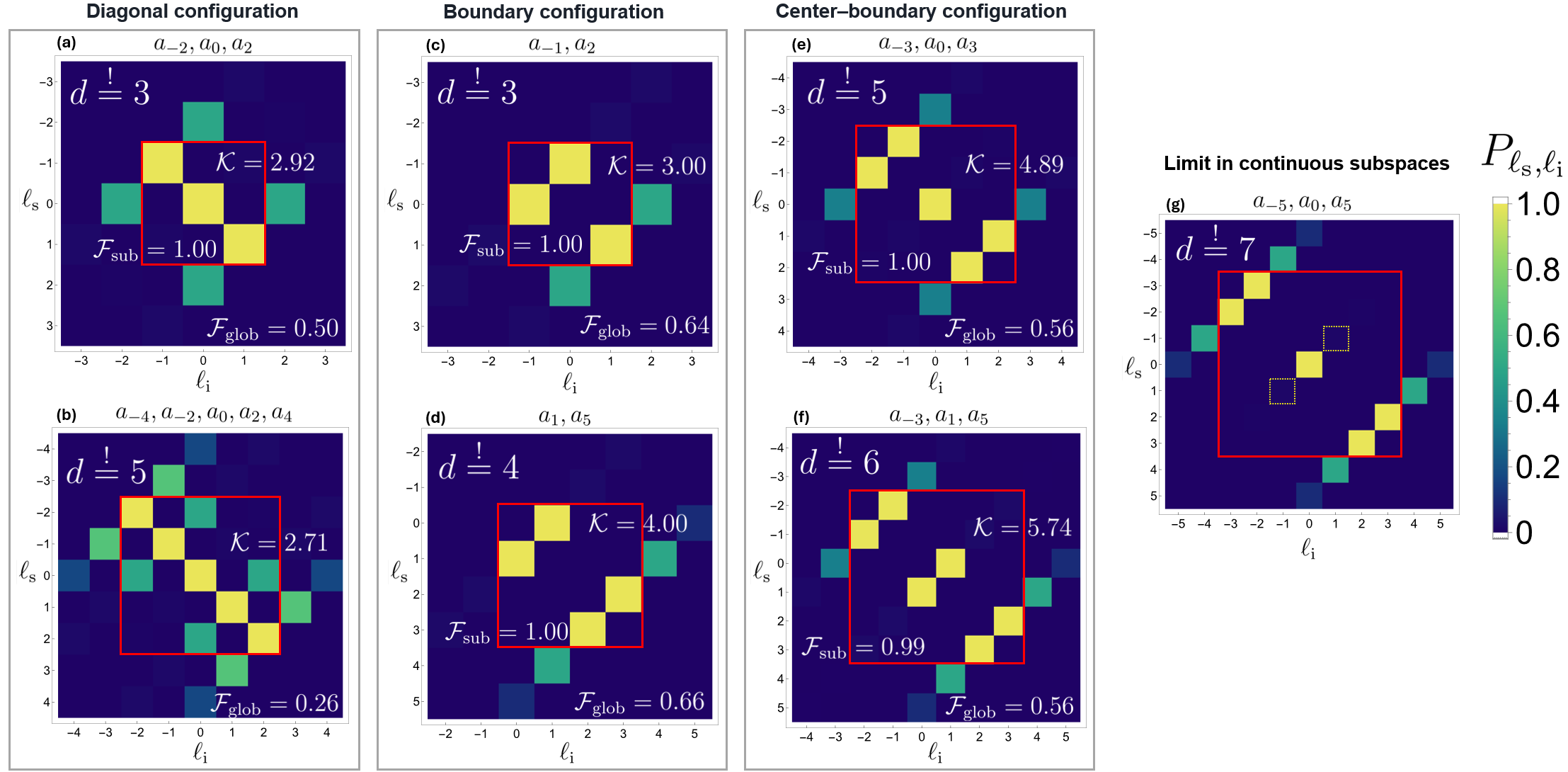}
    \caption{ Different methods to engineer MES within \textit{continuous} subspaces. The fidelities $\mathcal{F_{\text{sub}}}$ and $\mathcal{F_{\text{glob}}}$ quantify the overlap with the target MES within the selected subspace, marked in red, and within the full LG mode space, respectively. Populating the diagonal 
    $\ell_{\signal}=\ell_{\idler}$ works for $d=3$ in (a), but fails for larger dimensions due to non-vanishing mode contributions within the target subspace, as shown for $d=5$ in (b). Alternatively, by placing modes at or near the subspace boundary in the upper left or lower right in (c) and (d), unwanted contributions are suppressed. A hybrid method is shown in (e) and (f) by populating modes at the center of the subspace and at the boundary. Scaling to higher dimensionality $d$ requires equalizing modes neighboring the center modes, as indicated by the yellow dashed boxes in (g), which cannot effectively be done with pump beam shaping.}
    \label{figure02}
\end{figure*}

\subsection{Subspaces and maximal entanglement}
Although the OAM Hilbert space is formally infinite, the number of significantly populated modes in a practical setup is finite. We restrict the state preparation and measurement to selected $d$-dimensional OAM \textit{subspaces}. Treating the signal and idler photons as qudits, we define the $d$-dimensional single-photon subspace $\mathcal{S}_d = \text{span}\{ \ket{\ell_j}\}, j=1,...,d$. We further distinguish between \textit{continuous} OAM subspaces, with modes in a continuous interval of OAM mode numbers, and subspaces in a discontinuous range of OAM mode numbers, here called \textit{disassembled} OAM subspaces. For the biphotonic system, the associated two-photon subspace is $\mathcal{S}_{d \times d} = \mathcal{S}_{d} \otimes \mathcal{S}_{d}$ of dimension $d^2$.





By definition, a maximally entangled state in this subspace is a pure bipartite state for which tracing over either subsystem yields a maximally mixed state for the other \cite{horodecki2009quantum}. Here, we quantify the entanglement within the two-photon subspace $\mathcal{S}_{d \times d}$ using the Schmidt decomposition \cite{miatto2012spatial,miatto2012cartesian} and the Schmidt number $\mathcal{K}$ bounds the entanglement, ranging from $\mathcal{K}=1$ for a separable state to $\mathcal{K}=d$ for an MES \cite{bernecker24high}. For the states described by Eqs. \eqref{StateLGpump} and \eqref{Fullstate}, an MES in $\mathcal{S}_{d \times d}$ is formed by a coherent superposition of $d$ orthogonal mode pairs, each with an amplitude of $1/\sqrt{d}$, while the remaining $d^2-d$ mode amplitudes vanish. The condition of maximal entanglement $\mathcal{K}=d$ is satisfied only if both these requirements are met. Definitions based solely on equal amplitudes \cite{kovlakov2018quantum, baghdasaryan2020enhanced} are incomplete, as such states can result in reduced states that are not maximally mixed. In the OAM basis, the two conditions correspond to exactly one populated two-photon mode in each row and each column of the $d \times d$ subspace, with all $d$ amplitudes being equal. As this logic is very similar to the rule of filling a number in a Sudoku puzzle, we refer to this as the \textit{Sudoku condition}. The rest of the modes within the subspace must have negligibly small amplitudes. Since local unitary operations relate all MES within a given OAM subspace, it is sufficient to concentrate on a single MES for each dimension $d$.

Any population outside the target subspace decreases the effective photon-pair rate, as these components are projected out when measuring solely within the selected OAM modes. These contributions outside of the considered subspace are rarely thoroughly quantified in most works. Hence, we introduce the global fidelity $\mathcal{F}_{\text{glob}} = |\braket{\psi_{\text{tar}}|\psi}|^2$ that measures the overlap of the full LG state $\ket{\psi}$ and the maximally entangled target state $\ket{\psi_{\text{tar}}} \in \mathcal{S}_{d \times d}$. A value of $\mathcal{F}_{\text{glob}}=1$ indicates that the entire generated OAM spectrum perfectly matches the target state. We also define a subspace fidelity, where the full state $\ket{\psi}$ is restricted to the subspace modes. Thus, $\mathcal{F}_{\mathrm{sub}}=1$ corresponds to a maximally entangled state within $\mathcal{S}_{d\times d}$.

We note that we will search by design for conditional states that are only maximally entangled 
 after projection onto a selected OAM subspace for a given dimensionality. If the amplitudes of the selected target modes are not the dominant amplitudes of the full generated spectrum, the state preparation can be globally very inefficient and should be discarded. 
Figs. \ref{figureAppendix02neuW} and \ref{figureAppendix03neuW} in Appendix \ref{Appendix: Relative pump weights} illustrate such cases.





\section{Results}

A coherent superposition of appropriately weighted pump OAM modes provides a direct tool for satisfying the Sudoku condition within a selected $d$-dimensional OAM subspace.
Here, we investigate the relationship between the choice of target modes for signal and idler, the structure of the local OAM subspaces, and the OAM superposition of the pump beam.

\subsection{Dimensionality limit in continuous subspaces}
We first consider continuous subspaces of dimension $d$ and show that our construction of subspace-optimized MES with $\mathcal{K} \approx d$ can only be realized up to $d=6$. As discussed above, depending on the parity of $\ell_{\pump}$, we can have one central mode (even $\ell_{\pump}$) or two central modes (odd $\ell_{\pump}$) with maximum amplitude per anti-diagonal. We can recognize three different configurations to realize an MES for $d \leq 6$. 

The first configuration is to populate modes along the diagonal $\ell_{\signal} = \ell_{\idler}$ 
as shown in Fig.~\ref{figure02} (a) yielding the MES $\frac{1}{\sqrt{3}} \bigl( \ket{-1,-1} + \ket{0,0} + \ket{1,1} \bigr)$ in the subspace $\mathcal{S}_{3 \times 3}$ defined by $\mathcal{S}_3=\text{span} \{ \ket{-1}, \ket{0}, \ket{1} \}$. As a direct consequence of the non-vanishing modes adjacent to the target modes, we have $\mathcal{F}_{\text{glob}}<1$. This method fails for larger $d>3$: for example in $\mathcal{S}_5=\text{span} \{ \ket{-2},...,\ket{2} \} $, OAM conservation introduces additional contributions from $a_{\pm2}$ that reduce the Schmidt number, and the state is no longer maximally entangled, see Fig.~\ref{figure02} (b). The global fidelity in Fig.~\ref{figure02} (b) is further reduced compared to Fig.~\ref{figure02} (a) due to the increased number of terms in the pump superposition.

For $d=3$ and $d=4$, a boundary-maxima configuration can be used, in which the central maxima of the selected anti-diagonals are placed near the boundaries of the $d\times d$ OAM subspace. This arrangement shifts most unwanted neighboring contributions outside the selected mode space. For $\mathcal{S}_3$, the pump components $a_{-1}$ and $a_2$ yield the MES $\frac{1}{\sqrt{3}} \bigl( \ket{-1,0} + \ket{0,-1} + \ket{1,1} \bigr)$,  which achieves a higher global fidelity than the diagonal configuration, see Fig.~\ref{figure02} (c). The same principle extends to $d=4$, as shown in Fig.~\ref{figure02} (d).

\begin{figure*}
    \centering
    \includegraphics[width=0.83\linewidth]{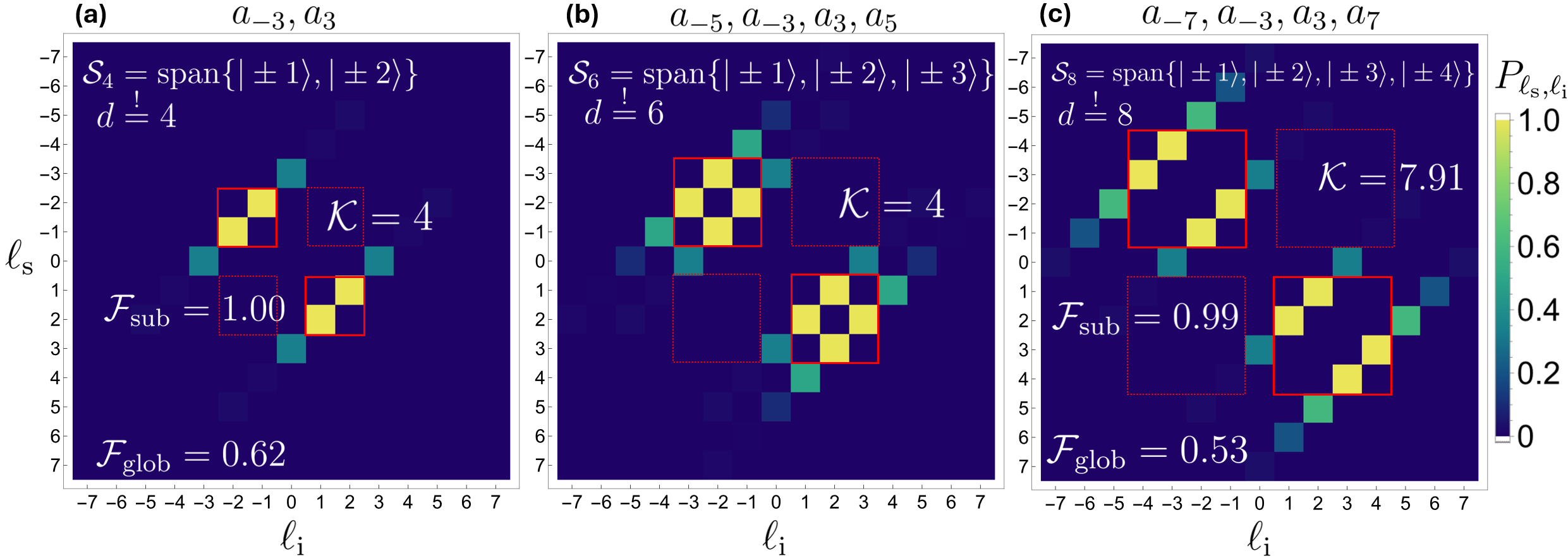}
    \caption{Pump engineering scheme in \textit{disassembled} subspaces. (a) The range of the available modes in a disassembled subspace is not continuous. The choice of pump modes determines which signal and idler modes can be selected by picking the maxima along the anti-diagonals. Compared to Fig. \ref{figure02} (e), we omit $a_0$ in the pump superposition, and the remaining pump modes generate four degenerate central amplitudes. We select only the OAM modes that support these amplitudes, which defines the disassembled subspace $\mathcal{S}_4$. (b) Not all pump configurations preserve maximal entanglement: $a_{\pm 5}$ introduces separable modes within the subspace $\mathcal{S}_6$, reducing the subspace fidelity. The state is not maximally entangled. (c) A sufficient difference in the OAM indices of the pump ensures well-separated two-photon modes that can form MESs. When the subspace is extended, e.g., by including modes $\ket{\pm3}$ and $\ket{\pm4}$, the appropriate choice of pump modes $a_{\pm 7}$ enables an MES for $d=8$ and excludes unwanted contributions.}
    \label{figure03}
\end{figure*}

For $d=5$ and $d=6$, the boundary configuration alone is no longer sufficient. Instead, degenerate modes near the boundary must be combined with modes in the subspace center. Such center modes can only be prepared with pump weights $a_0$ and $a_{\pm1}$, since all other pump modes introduce undesired, adjacent modes inside the subspace. This enables the only viable MES in $\mathcal{S}_{5 \times 5}$ shown in Fig.~\ref{figure02} (e). For continuous $\mathcal{S}_{6 \times 6}$ subspaces we can use $a_{\pm1}$ for the center modes. An example is shown in Fig.~\ref{figure02} (f) with $\mathcal{S}_6=\text{span} \{ \ket{-2}, ..., \ket{3} \}$. 


For $d \geq 7$, however, continuous subspaces will fail: exactly matching only a few required amplitudes to those of the center modes, like the two modes indicated by the yellow dashed square in Fig.~\ref{figure02} (g), cannot be efficiently achieved by tuning the pump alone \cite{bernecker2025engineering}. 

Appendix \ref{Appendix: Comparison with large pump waists} discusses matching different amplitudes along the same anti-diagonal $\ell_{\pump}=0$ for large pump beam waist (approaching the plane-wave-limit). Including higher-order radial pump modes could enable equal amplitudes of such modes as well, but can significantly reduce the global fidelity due to increased radial correlations. For example, Fig. \ref{figureAppendix03neuW} (b) in the Appendix shows a MES for $d=7$ with  Schmidt number $K=6.91$. Preparing this state, however, comes at the cost of a lower global fidelity and the loss of the analytical ease of calculating the pump spectrum. This motivates the use of disassembled subspaces.




\subsection{MES preparation based on disassembled subspaces}

We first illustrate the idea of a disassembled subspace with a modified configuration from the continuous subspace of Fig.~\ref{figure02} (e): removing the pump component $a_0$ gives a spectrum shown in Fig.~\ref{figure03} (a).
The remaining pump weights $a_{\pm 3}$ generate four degenerate central amplitudes. Instead of keeping the continuous interval from before, we select only the OAM modes that support these amplitudes, defining the subspace $\mathcal{S}_4=\mathrm{span} \bigl\{ \ket{-2},\ket{-1},\ket{1},\ket{2} \bigl\}$. Within this disassembled subspace, the generated state $ \ket{\psi}= \frac{1}{2} \bigl( \ket{-2,-1} + \ket{-1,-2} + \ket{1,2} + \ket{2,1} \bigl)$ is a four-dimensional MES. This example demonstrates the basic role of disassembled subspaces: modes that are not needed for the target state are excluded, so that unwanted contributions do not enter the effective $d\times d$ Hilbert space. The global fidelity is $\mathcal{F}_{\text{glob}}=0.62$, slightly lower than in the continuous-subspace example (Fig. \ref{figure02} (d)), as our choice of pump modes here increases the number of undesired modes.

Next, we show that this approach can be extended beyond the dimensionality limit with $(d>6$) for MESs in a continuous subspace. As an example, we consider the dimension $d=8$. The subspace from Fig.~\ref{figure03} (a) can be extended stepwise by adding suitable pump terms. A natural first attempt is to use four odd pump components and equalize their central amplitudes. Yet, amplitude equalization alone is not sufficient. As shown in Fig.~\ref{figure03} (b), choosing pump modes that are too close to each other ($a_{\pm5}$) can produce eight equally weighted amplitudes, but the 
Sudoku condition for an eight-dimensional MES is violated. Consequently, the corresponding mode pairs are not independent, and signal and idler modes are reused within the subspace. The subspace modes form a $6 \times 6$ subspace, but the state is not maximally entangled, $\mathcal{K}=4 \neq 6$.

This problem is avoided by increasing the separation between the anti-diagonals, i.e., the difference in pump OAM indices. In particular, replacing the additional pump components by $a_{\pm7}$ yields new central mode pairs that do not overlap with those generated by $a_{\pm3}$. The resulting target modes define the disassembled subspace $ \mathcal{S}_8=\mathrm{span} \bigl\{ \ket{-4},\ket{-3},\ket{-2},\ket{-1}, \ket{1},\ket{2},\ket{3},\ket{4} \bigl\}$, within which the eight equal amplitudes form a 
eight-dimensional MES, cf. Fig.~\ref{figure03} (c).

This illustrates the central principle of the approach: the OAM subspace and the choice of the OAM pump mode weights are closely related. High-dimensional MESs can be constructed by introducing sufficient separation between the pump modes, equalizing the desired central amplitudes through their relative weights, and selecting a subspace that retains these target modes while excluding unwanted neighboring contributions. In general, several valid pump-subspace combinations may exist for a given dimension. 

When comparing Fig. \ref{figure01} (d) to Fig. \ref{figure02} (a) and (c), we recognize that multiple possibilities for $d=3$ exist to choose pump modes or subspace modes. 
Higher OAM indices for the pump modes can 
equivalently enable conditioned maximally entangled qutrits, but the number of non-vanishing modes adjacent to the produced target modes increases. The same holds for the construction of MESs in disassembled states. 
The following design criteria
for suitable MESs and subspaces can be summarized as:
\begin{enumerate}
    \item Fulfill the Sudoku condition (one target mode per column and row, all with equal amplitudes). The OAM spectrum in the conditioned subspace represents a $d \times d$ MES with Schmidt number $\mathcal{K} \approx  d$. In the broadband regime, we have to make sure that all target modes have the same spectral dependence to ensure spatial purity.
    \item The amplitudes within the subspace are the global maxima of all amplitudes. 
    Prevent pump modes that generate the most probable modes for signal and idler outside the target subspace. 
    \item Maximize the global fidelity via waist optimization to enhance the population within the target subspace and to reduce mode leakage contributions outside the subspace.
\end{enumerate}


\begin{figure}[b]
    \includegraphics[width=0.85\linewidth]{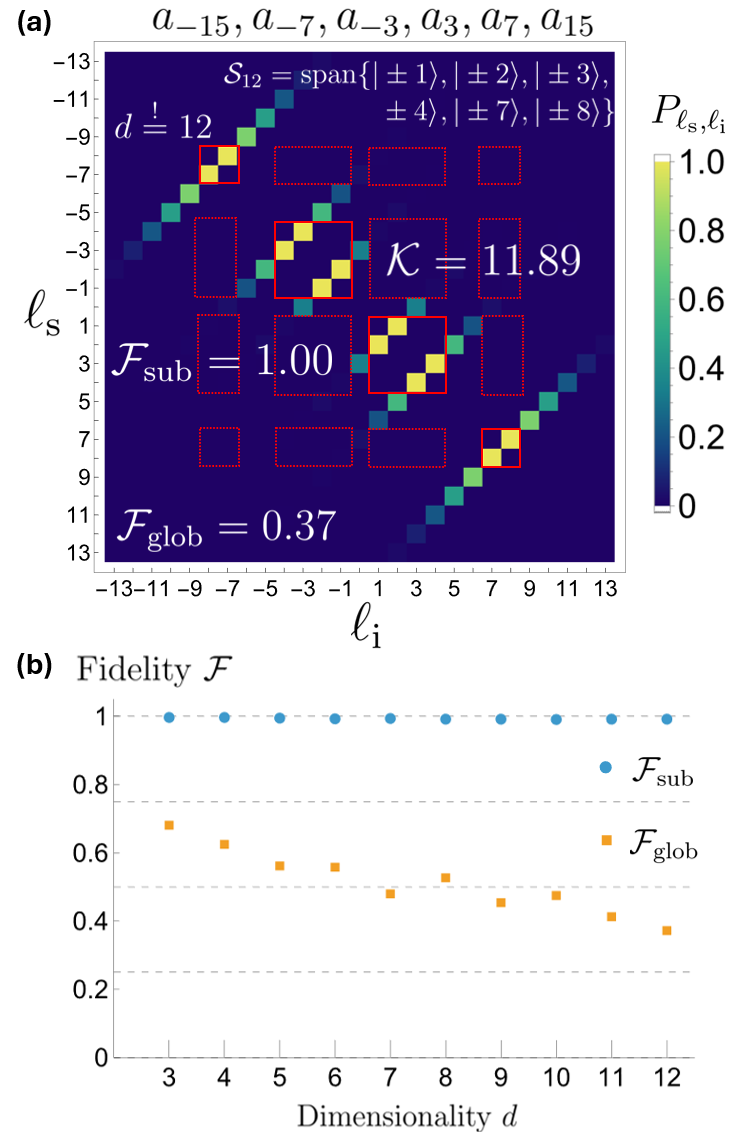}
    \caption{Joint pump-beam and modal space engineering based on disassembled subspaces extended to higher dimensions. (a) Six sufficiently separated pump OAM components are used to construct a $d=12$ MES. Compared to Fig. \ref{figure03} (c), the subspace requires modes with two additional gaps within the selected subspace range. (b) A summary of the subspace and global fidelities obtained with our approach as functions of the MES dimensionality up to $d=12$.}
    \label{figure04}
\end{figure}

To maximize the global fidelity, one should minimize both the number of pump components and the spiral bandwidth of each component. The latter can be achieved in the SSMR. Odd pump OAM modes are particularly useful because each of them contributes two degenerate target mode pairs, enabling larger dimensionalities with fewer pump modes.

Fig. \ref{figure04} (a) shows how our approach can be readily extended to larger dimensions. For example, we make use of $a_{\pm15}$ and find a $12 \times 12$ MES if subspace modes $\{ \ket{\pm7}, \ket{\pm8} \}$ are included, and $\{ \ket{\pm5}, \ket{\pm6} \}$ are excluded. A further increase in dimensionality requires even larger separations than that between $\pm 7$ and $\pm 15$. 

Larger $\ell_{\pump}$ results in a wider bandwidth with non-negligible modes and therefore smaller global fidelities. Therefore, the general trend is that increasing $d$ of the MES results in a decrease in global fidelity. Figure~\ref{figure04} (b) compares the subspace and global fidelities obtained with our approach as functions of the dimensionality up to $d=12$. More details on how the figure is obtained, and which pump and subspace modes are chosen, can be found in Appendix \ref{Appendix: Detailed description of Figure 4 (b)}.

Karan et al.
\cite{karan2023postselection} introduced a postselection-free
approach based on higher-order radial pump modes, enabling
the engineering of approximately uniform OAM spectra
while leaving the radial degree of freedom unresolved.
Although this approach avoids the losses associated with
selecting only the lowest radial order, equalized OAM
populations do not necessarily imply a pure maximally
entangled OAM state: correlations with unresolved radial
or spectral degrees of freedom can reduce coherence
between OAM components. In our approach, purity is an
explicit design requirement. We select a defined
LG-mode subspace whose target mode pairs share the
same spectral dependence, ensuring spatial--spectral
separability of the target MES.  A possible extension of our approach to include higher radial modes is discussed in Appendix \ref{Appendix: Include higher radial indices of signal and idler}.

\section{Conclusion}
In this paper, we introduced a universal pump engineering approach for the design of high-dimensional MESs via SPDC. Our analysis demonstrates that the joint optimization of the spatial structure of the pump and the careful design of the effective OAM subspace enable the preparation of MESs. For continuous OAM subspaces, the finite spiral bandwidth imposes a dimensionality limit of $d\leq6$, since the construction of higher-dimensional subspaces inevitably leads to undesired, non-negligible mode contributions within the target space. To overcome this limitation, we introduced the concept of disassembled OAM subspaces. By engineering the OAM pump modes and selecting subspace modes, this approach isolates the desired maximal amplitudes while systematically eliminating unwanted contributions arising from the spiral bandwidth. Ultimately, our combined pump–modal space engineering provides a practical yet extendable approach to generating high-dimensional MESs with, in principle, arbitrary dimensionality.


\appendix

\section{Details on the biphoton state}
\label{Appenndix: Details on the biphoton state}
SPDC, as a second-order nonlinear process, is inherently probabilistic. We assume a classical, undepleted pump beam driving the SPDC process in the low-gain regime. Here, one pump photon splits into a pair of signal and idler photons when interacting with a second-order nonlinear material. The photons propagate quasi-collinear along $\bm{k}_j= \bm{q}_j + k_{j,z} \; \mathrm{e}_z$ with $|\bm{q}| \ll |\bm{k}|:=k$ for $j= \pump, \signal, \idler$, respectively. In the paraxial regime, the down-converted photons of frequency $\omega_j$, transverse momentum $\bm{q}_j$, and polarization $\sigma_j$ are generated by creation operators $\hat{a}^{\dagger}$ acting on the vacuum states as
\begin{align*} 
    \ket{\psi} \propto \sum_{\sigma_{\signal},\sigma_{\idler}} & \int d^2\bm{q}_{\signal} \:d\omega_{\signal} d^2\bm{q}_{\idler} \:d\omega_{\idler} \; \Phi(\bm{q}_{\signal},\omega_{\signal},\bm{q}_{\idler},\omega_{\idler}) \nonumber \\
    & \times \biggl( \hat{a}_{\sigma_{\signal}}^{\dagger} (\bm{q}_{\signal},\omega_{\signal}) \: \otimes \hat{a}_{\sigma_{\idler}}^{\dagger}(\bm{q}_{\idler},\omega_{\idler}) \biggl) \ket{\text{vac}}.
\end{align*}
The biphoton mode function $\Phi(\bm{q}_{\signal},\omega_{\signal},\bm{q}_{\idler},\omega_{\idler})$ reads 
\begin{align*}
   \Phi(&\bm{q}_{\signal},\bm{q}_{\idler},\omega_{\signal}, \omega_{\idler}) = V_{\pump}(\bm{q}_{\signal}+\bm{q}_{\idler}) \: S_{\pump}(\omega_{\signal} + \omega_{\idler}) \\
   & \times \int_{-L/2}^{L/2} dz \: \chi^{(2)}(z) \: \mathrm{e}^{i \bigl( k_{\pump,z}(\bm{q}_{\signal}+\bm{q}_{\idler}) - k_{\signal,z}(\bm{q}_{\signal})  - k_{\idler,z}(\bm{q}_{\idler})  \bigl) z},
\end{align*} 
where the variable nonlinear susceptibility profile $\chi^{(2)}(z)$ represents either a periodically poled crystal \cite{torres2004quasi} or a custom-poled structure \cite{dosseva2016shaping}. $V_{\pump}$ and $S_{\pump}$ are the spatial and the spectral distribution of the pump beam, respectively. The biphoton mode function encodes the spatial and spectral properties of the photon pairs and is, in general, non-separable. Transverse momentum conservation $\bm{q}_{\pump}=\bm{q}_{\signal}+\bm{q}_{\idler}$ is justified when the transverse crystal dimensions are much larger than the beam waists. There are no restrictions to the longitudinal direction, i.e., the length $L$ of the crystal. The phase-matching conditions determine the propagation directions of the photons, the allowed frequency range, and the permitted polarization combinations. We can omit the polarization index $\sigma_j$ implicitly by specifying the phase matching as type-0, type-I, or type-II SPDC.




\begin{figure*}
    \centering
    \includegraphics[width=0.75\linewidth]{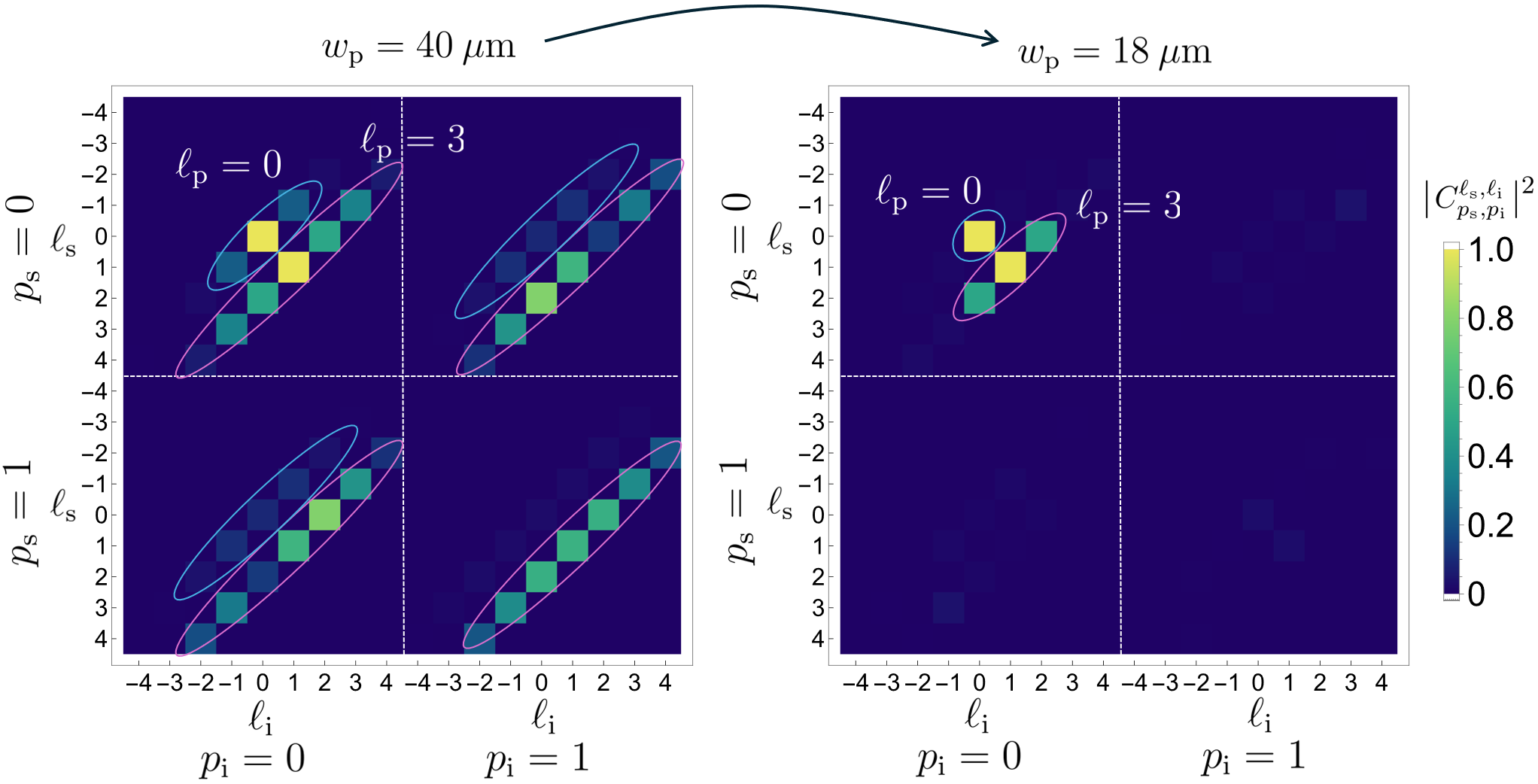}
    \caption{OAM spectra compared for different pump beam waists. OAM spectra pumped with a superposition of OAM indices $\ell_{\pump}=0,3$ like in Fig. 1 of the main text, but radial indices $p_{\signal}=p_{\idler}=0,1$ are included. In the SSMR, most mode distribution is concentrated for the lowest radial order, minimizing the extent of the spiral bandwidth. The regime can be achieved for a waist ratio $\gamma=\frac{w_{\pump}}{w_{\signal,\idler}}=\frac{1}{\sqrt{2}}$. Here, we set $w_{\signal,\idler}=25 \; \mu$m.}
    \label{figureAppendix01neuW}
\end{figure*} 

An important aspect of SPDC-based photon sources is the spectral-spatial coupling. Since we are primarily interested in the spatial domain (transverse momentum space), the narrowband (or quasi-monochromatic) regime, achieved experimentally using narrowband spectral filters, effectively projects the signal and idler onto the central frequencies $\omega_{\signal,0}$ and $\omega_{\idler, 0}$ satisfying $\omega_{\pump,0}=\omega_{\signal,0}+\omega_{\idler,0}$ \cite{sevilla2024spectral}. Consequently, all subsequent expressions are evaluated at these fixed frequencies. For a broadband spectral description we use $\omega=\omega_0+\Omega$, where the frequency deviation $\Omega$ is much smaller than the central frequency $\: \Omega \ll \omega_0$. Then the Fresnel approximation yields
\begin{small}
\begin{align*}
  k_z  =k(\Omega)\sqrt{1-\frac{|\bm{q}|^2}{k(\Omega)^2}} = k(\omega_0)+\frac{\Omega}{u(\omega_0)}+\frac{G(\omega_0) \: \Omega^2}{2}-\frac{|\bm{q}|^2}{2k(\omega_0)},
  \label{fresnelapprox2}
\end{align*}
\end{small}
with the group velocity $u$ and the group velocity dispersion $G$. For a phase-matched crystal satisfying $k_{\pump}-k_{\signal}-k_{\idler}=0$, the remaining phase mismatch is given by
\begin{align*}
    \Delta k_z = \frac{| \bm{q}_{\signal}|^2 }{ 2k_{\signal}(\omega_{\signal})}  + \frac{ |\bm{q}_{\idler}|^2 }{ 2k_{\idler}(\omega_{\idler})} - \frac{|\bm{q}_{\signal}+\bm{q}_{\idler}|^2 }{ 2k_{\pump}(\omega_{\pump})} + \Delta_{\Omega}
\end{align*}
with
\begin{small}
\begin{align*}
   \Delta_{\Omega}=  \frac{\Omega_{\signal}+\Omega_{\idler}}{u_{g,\pump}} - \frac{\Omega_{\signal}}{u_{g,\signal}} - \frac{\Omega_{\idler}}{u_{g,\idler}} + \frac{1}{2} \Bigl( G_{\pump} (\Omega_{\signal}+\Omega_{\idler})^2 - G_{\signal} \Omega_{\signal}^2 -  G_{\idler} \Omega_{\idler}^2  \Bigl)
\end{align*}
\end{small}
only being present in the broadband regime. Unless stated otherwise, our analysis uses the following baseline parameters: a 405 nm continuous-wave (cw) laser pumping a 15 mm ppKTP crystal for type-II phase matching, producing photons spectrally centered at 810 nm. Importantly, our method is independent of these specific parameters and is readily adaptable to other configurations.

\section{Expansion amplitude of the full LG basis}
\label{Appenndix:Expansion amplitude of the full LG basis}
Transverse LG modes also carry the radial index $p$ in addition to the OAM index $\ell$. Due to the conservation of OAM, the azimuthal index $\ell$ plays a central role in the transverse spatial description of SPDC and remains the primary focus of this work. Consequently, we adopted a simplified notation for the LG modes in the main text, omitting the radial index $p$. Here, we introduce a notation that captures the complete transverse description of the two-photon state and explain how it relates to the notation used in the main text. When calculating the global fidelity of the full state, the radial description cannot be neglected. The general LG decomposition of the biphoton state is
\begin{equation}
    \ket{\psi} = \sum_{p_{\signal},\ell_{\signal}}  \sum_{p_{\idler},\ell_{\idler}} C_{p_{\signal},p_{\idler}}^{\ell_{\signal},\ell_{\idler}} \; \biggl( \ket{p_{\signal},\ell_{\signal}}_{\signal} \otimes \ket{p_{\idler},\ell_{\idler}}_{\idler} \biggl ) ,
    \label{Fullstatewithp}
\end{equation}
where the spatial part is expanded into the complete set of orthogonal LG modes $\ket{p,\ell} = \int d\bm{q} \:  \text{LG}(p,\ell,w) \: \hat{a}^{\dagger}(\bm{q}) \ket{\text{vac}}$, that are eigenstates of the OAM operator $\hat{L}\ket{p,\ell} = \ell \hbar \ket{p,\ell}$, $\ell \in \mathbb{Z}$ and where $w$ is the waist of the LG mode. The full expression for the LG modes in momentum space is
\begin{eqnarray*}
     \mathrm{LG}^{\ell}_{p}(\bm{q},w) = \sqrt{\frac{w^2 p!}{2\pi(p + |\ell|)!}} (-1)^{p+\frac{\ell}{2}} \left( \frac{w |\bm{q}|}{\sqrt{2} } \right)^{|\ell|} \\
     \times \mathrm{e}^{-\frac{|\bm{q}|^2 w^2}{4}} \; \mathrm{L}_{p}^{|\ell |}\left(\frac{|\bm{q}|^2 w^2}{2}\right) \mathrm{e}^{i \ell \mathrm{Arg}(\bm{q}) }.
\label{LGexpression}
\end{eqnarray*}
The experimental parameters used to tune the expansion amplitudes can be grouped into three categories: the transverse spatial distribution of the pump beam,
the structure of the nonlinear crystal, and the geometry of the measurement setup. When inserting $V_{\pump}(\bm{q}_{\signal}+\bm{q}_{\idler}) =  \mathrm{LG}^{\ell}_{p}(\bm{q}_{\signal}+\bm{q}_{\idler},w) $, we obtain (in the narrowband regime)
\begin{widetext}
\begin{eqnarray*}
     && C^{\ell_{\signal},\ell_{\idler}}_{p_{\signal},p_{\idler}} \bigl( \underbrace{p_{\pump},\ell_{\pump},w_{\pump}}_{\mathrm{pump}}; \underbrace{\chi^{(2)},L}_{\mathrm{crystal}}; \underbrace{w_{\signal},w_{\idler}}_{\mathrm{measurement}} \bigl) = \Bigl( \bra{{p_{\signal},\ell_{\signal}}} \otimes  \bra{{p_{\idler},\ell_{\idler}}} \Bigl) \ket{\psi} \\
    &&= \mathcal{N} \int_{-\frac{L}{2}}^{\frac{L}{2}} dz \iint d\bm{q}_{\signal} \; d\bm{q}_{\idler} \; \;  \chi^{(2)}(z)  \; \mathrm{e}^{ i \Delta k_z(\bm{q}_{\signal},\bm{q}_{\idler}) z} \; 
   \mathrm{LG}_{p_{\mathrm{p}}}^{\ell_{\pump}}(\bm{q}_{\signal}+\bm{q}_{\idler}, w_{\pump}) \left[ \mathrm{LG}_{p_{\signal}}^{\ell_{\signal}}(\bm{q}_{\signal},w_{\signal}) \right]^* \left[ \mathrm{LG}_{p_{\idler}}^{\ell_{\idler}} (\bm{q}_{\idler},w_{\idler}) \right]^* 
    \propto \delta_{\ell_{\signal}+\ell_{\idler},\ell_{\pump}}. 
    \label{Amplitudeexpression}
\end{eqnarray*}
\end{widetext}
The collection waists for signal and idler $w_{\signal}, w_{\idler}$ are determined by the experimental setup and projective detection scheme. The Kronecker delta expresses the well-known OAM conservation and the resulting OAM anti-correlations between signal and idler photons $\ell_{\idler}=\ell_{\pump}-\ell_{\signal}$. 

\begin{figure*}
    \centering
    \includegraphics[width=0.8\linewidth]{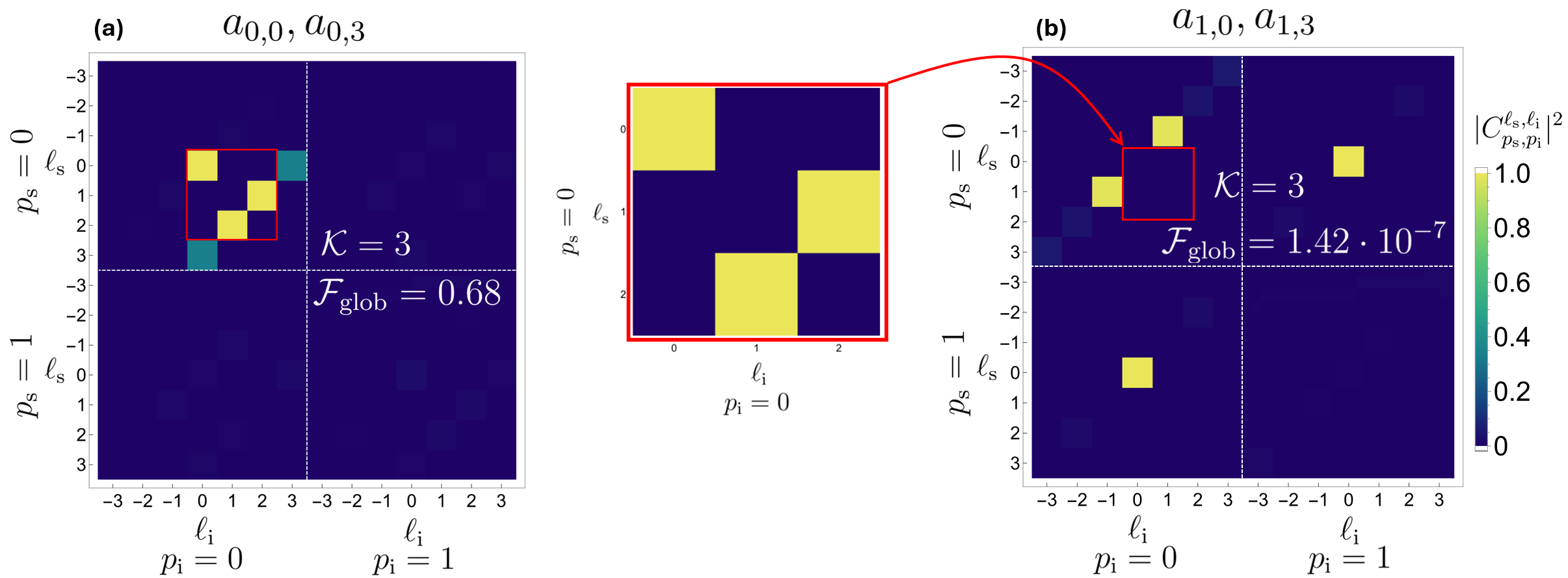}
    \caption{OAM spectra pumped with superposition of OAM indices $\ell_{\pump}=0,3$, but different radial indices of the pump beam compared. (a) Fig. 1 with $p_{\pump}=0$ reconstructed. Within the subspace we find $\mathcal{K}=3$, that corresponds to $\mathcal{F}_{\text{sub}}=1$ for the target MES. The global fidelity is $\mathcal{F}_{\text{glob}} = 0.68$. (b) For $p_{\pump}=1$, we locally generate the same MES, optimized in the subspace. However, the fidelity is practically zero since these are only small contributions to the global state. This illustrates that subspace optimization is only efficient using the maxima of the anti-diagonals as target modes.}
    \label{figureAppendix02neuW}
\end{figure*} 

This description of SPDC applies to a wide range of modern SPDC experiments with entangled photons without the need for additional approximations. For both the narrowband and broadband regimes, analytical formulas and compact closed-form expressions for the biphoton state can be derived, as shown in detail in \cite{baghdasaryan2022generalized, bernecker24high}. The main idea of the derivation is based on expressing the Laguerre polynomials as a sum,
\begin{equation*}
         \mathrm{LG}^{\ell}_{p}(\bm{q},w) = \mathrm{e}^{-\frac{|\bm{q}|^2 w^2}{4}} \; \mathrm{e}^{i \ell \mathrm{Arg}(\bm{q}) } \sum_{u=0}^p T_u^{p,\ell}(w) \; | \bm{q}|^{2u + |\ell|}
         \label{LGwithsum}
\end{equation*}
with the polynomial expansion coefficient
\begin{align*}
   T_u^{p,\ell}(w) =  \sqrt{\frac{p!\,(p+|\ell|)!}{\pi}} \;  \biggr(\frac{ w}{\sqrt{2}}\biggl)^{2u+|\ell|+1} \frac{(-1)^{p+u} \; i^{\ell}}{(p - u)! \; (|\ell| + u)! \; u!}, 
\end{align*}
which allows deriving analytical expressions in polar coordinates. If the nonlinear crystal is instead pumped by an arbitrary transverse field expressed by LG modes, 
\begin{align*}
V_{\pump} = \sum_{p_{\pump},\ell_{\pump}}   a_{p_{\pump},\ell_{\pump}}^{} \: \text{LG}_{}(p{_{\pump}}, \ell{_{\pump}}, w_{\pump}),
\end{align*}
the amplitudes are updated according to the linearity of the SPDC amplitude in the pump field to
\begin{eqnarray*}
{C^{\prime}}^{\ell_{\signal},\ell_{\idler}}_{p_{\signal},p_{\idler}}  (  a_{p_{\pump},\ell_{\pump}}^{}, w_{\pump}) \propto \sum_{p_{\pump},\ell_{\pump}}  a_{p_{\pump},\ell_{\pump}}^{} \; C^{\ell_{\signal},\ell_{\idler}}_{p_{\signal},p_{\idler}} ( p_{\pump},\ell_{\pump}, w_{\pump}).
\end{eqnarray*}

The coefficients are normalized according to $ \sum_{p_{\pump},\ell_{\pump}}  |a_{p_{\pump},\ell_{\pump}}|^{2}=1$.
Although the $p$ dependence was analyzed in great detail, there is no corresponding conservation rule. For experimental convenience, the $p$ index is often omitted, and therefore it has often been referred to as the "forgotten quantum number" \cite{plick2015physical}. In the single Schmidt mode regime (SSMR), the correlations of higher-order radial modes $p>0$ of the OAM spectrum nearly vanish for $p_{\pump}=0$, and the dominant contribution to the state is concentrated in the mode $p_{\signal}=p_{\idler}=0$. Figure \ref{figureAppendix01neuW} shows the same distribution as Fig. \ref{figure01}(b) and (c) of the main text, illustrating this behavior. The SSMR is achieved when the Rayleigh range of the pump beam is equal to half the crystal length and the ratio between the pump and collection waists is set to $\gamma=\frac{w_{\pump}}{w_{\signal}}=\frac{w_{\pump}}{w_{\idler}}=\frac{1}{\sqrt{2}}$, as pointed out in \cite{kovlakov2017spatial, kovlakov2018quantum}. By optimizing the down-conversion regime for a minimal spiral bandwidth, we can maximize the global fidelity and justify our choice of target modes with $p=0$.

Inserting the Kronecker delta into Eq. \eqref{Fullstatewithp} and setting the radial indices of the signal and idler photons to be zero, such that $\ket{0,\ell_{\signal}}_{\signal}:=\ket{\ell_{\signal}}_{\signal}$ and $\ket{\ell_{\idler}}_{\idler}:=\ket{0,\ell_{\idler}}_{\idler}=\ket{0,\ell_{\pump}-\ell_{\signal}}_{\idler}$ leads to expression Eq. (1) from the main text if $C_{0,0}^{\ell_{\signal},\ell_{\idler}}  \equiv C_{0,0}^{\ell_{\signal},\ell_{\pump}-\ell_{\signal}}:=C_{\ell_{\signal}}$ is defined. Additionally, by writing $a_{0,\ell_{\pump}}^{}:=a_{\ell_{\pump}}^{}$ we recover Eq. (2) from the main text. To calculate the global fidelity, we use the full state, including higher radial orders, to accurately quantify how much of the target state population matches the experimentally expected spectra.


\section{Relative pump weights}
\label{Appendix: Relative pump weights}
The advantage of our proposed method is that it determines the relative pump weights using only the coincidence amplitudes $C^{\ell_{\signal},\ell_{\idler}}_{p_{\signal},p_{\idler}} (p_{\pump},\ell_{\pump}, w_{\pump})$. Here, we provide additional details on how to calculate the pump weights $a_{\ell_{\pump}}$ and explain why including higher radial pump modes is disadvantageous. As an example, we reconstruct the MES shown in Fig. 1 of the main text. The target MES in OAM  $\frac{1}{\sqrt{3}} \bigl( \ket{0,0} + \ket{1,2} + \ket{2,1}  \bigl)$ (for radial indices $p_{\signal}=p_{\idler}=0$) within the continuous subspace $\mathcal{S}_3=\text{span} \{ \ket{0}, \ket{1}, \ket{2} \}$ can be written as a matrix of coefficients. By assuming the same value of $w_{\pump}$ for all pump modes, we omit the waist dependence. This gives a coefficient matrix
\begin{widetext}
    \begin{equation*}
\begin{pmatrix}
C^{0,0}_{0,0}& C^{0,1}_{0,0} &  C_{0,0}^{0,2}  \\
C^{1,0}_{0,0} &  C^{1,1}_{0,0} & C_{0,0}^{1,2}   \\
 C^{2,0}_{0,0} & C^{2,1}_{0,0} & C_{0,0}^{2,2}  \\
\end{pmatrix} \longrightarrow
\begin{pmatrix}
a_{0,0} \; C^{0,0}_{0,0}(0,0) & a_{0,1} \; C^{0,1}_{0,0}(0,1) & a_{0,2} \; C_{0,0}^{0,2}(0,2)  \\
a_{0,1} \; C^{1,0}_{0,0}(0,1) & a_{0,2} \; C^{1,1}_{0,0}(0,2) & a_{0,3} \; C_{0,0}^{1,2}(0,3)   \\
a_{0,2} \; C^{2,0}_{0,0}(0,2) & a_{0,3} \; C^{2,1}_{0,0}(0,3) & a_{0,4} \;C_{0,0}^{2,2}(0,4) \\
\end{pmatrix}
\overset{!}{=} \frac{1}{\sqrt{3}} \begin{pmatrix}
1 & 0 & 0 \\
0 & 0 & 1 \\
0 & 1 & 0 \\
\end{pmatrix}. \\
\end{equation*}
\end{widetext}
After the arrow, we introduced a general pump superposition of OAM modes (with $p_{\pump}=0$) relevant in the subspace: $V_{\pump}(p_{\pump},\ell_{\pump})= a_{0,0} \: \text{LG}(0,0) + a_{0,1} \: \text{LG}(0,1) + a_{0,2} \: \text{LG}(0,2) + a_{0,3} \: \text{LG}(0,3) + a_{0,4} \: \text{LG}(0,4)$. Here, we assume that the projected state is normalized within the target subspace, so that the target amplitudes are $1/\sqrt{3}$.

The choice $a_{0,1}=a_{0,2}=a_{0,4}=0$ is immediate. For the other pump weights, we find $a_{0,0} = \bigl(\sqrt{3} \: C^{0,0}_{0,0}(0,0) \bigl)^{-1}$ as well as $a_{0,3} =  \bigl(\sqrt{3} \: C^{2,1}_{0,0} (0,3) \bigl)^{-1}$.  The degeneracy condition $k_{\pump}/2 = k_{\signal} = k_{\idler}$ along with $w_{\signal}=w_{\idler}$ ensures $C_{0,0}^{1,2}(0,3)=C^{2,1}_{0,0}(0,3)$, leading to the desired state. The corresponding OAM spectrum is again shown in Fig. \ref{figureAppendix02neuW} (a). Since we are in the SSMR, the contributions are predominantly concentrated within the $p_{\signal}=p_{\idler}=0$ sector.

We might expect that the same procedure can be extended to a pump superposition including modes with radial index $p_{\pump}=1$. It is then important to emphasize which pump mode gives rise to which two-photon amplitude. Although we obtain a similar system of equations,
\begin{widetext}
    \begin{equation*}
\begin{pmatrix}
a_{1,0} \; C^{0,0}_{0,0}(1,0) & a_{1,1} \; C^{0,1}_{0,0}(1,1) & a_{1,2} \; C_{0,0}^{0,2}(1,2)  \\
a_{1,1} \; C^{1,0}_{0,0}(1,1) & a_{1,2} \; C^{1,1}_{0,0}(1,2) & a_{1,3} \; C_{0,0}^{1,2}(1,3)   \\
a_{1,2} \; C^{2,0}_{0,0}(1,2) & a_{1,3} \; C^{2,1}_{0,0}(1,3) & a_{1,4} \;C_{0,0}^{2,2}(1,4) \\
\end{pmatrix}
\overset{!}{=} \frac{1}{\sqrt{3}} \begin{pmatrix}
1 & 0 & 0 \\
0 & 0 & 1 \\
0 & 1 & 0 \\
\end{pmatrix},
\end{equation*}
\end{widetext}
and find equivalent solutions $a_{1,1}=a_{1,2}=a_{1,4}=0, a_{1,0} = \bigl(\sqrt{3} \: C^{0,0}_{0,0} (1,0)\bigl)^{-1}, a_{1,3} =  \bigl(\sqrt{3} \: C^{2,1}_{0,0}(1,3) \bigl)^{-1}$, we observe that the fidelity is practically zero, as shown in Fig. \ref{figureAppendix02neuW} (b). The reason is that we are optimizing only within a restricted subspace. Within the target subspace, the requirement of equalized amplitudes is fulfilled, as apparent from the inset. Globally, since $C^{0,0}_{0,0}(1,0) \ll C^{-1,1}_{0,0}(1,0)$, the resulting weights lead to $\frac{1}{\sqrt{3}}\frac{C^{-1,1}_{0,0}(1,0)}{C^{0,0}_{0,0}(1,0)} \gg \frac{1}{\sqrt{3}}$. Consequently, the above choice results in an inefficient state preparation. As discussed in the main text, it is therefore preferable to restrict to those modes with maximal amplitudes generated by the pump and to use them as target modes.

\begin{figure*}
    \centering{
  }  \includegraphics[width=0.95\linewidth]{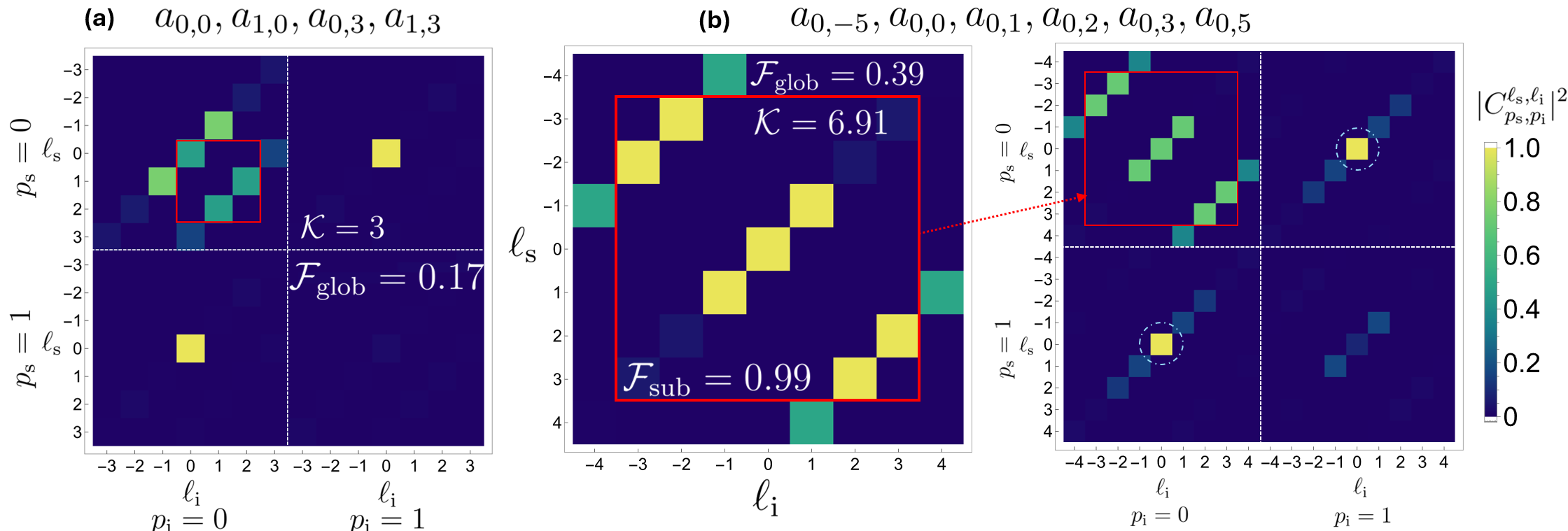}
    \caption{OAM spectra pumped with a superposition of mixed OAM and radial indices. (a) OAM spectrum for a spatial pump distribution with OAM modes $\ell_{\pump}=0,3$ and $p_{\pump}=0,1$. Similar to Fig. \ref{figureAppendix02neuW}, the maxima may lie outside of the intended subspace, resulting in a small global fidelity. (b) The same pump distribution as shown in Fig. 2 (g) of the main text, with additional higher radial modes $p_{\pump}=1,2,3$ for $\ell_{\pump}=0$ to overcome the continuous subspace limit. Now, within $\mathcal{S}_{7\times7}$ with $\mathcal{S}_7=\text{span}\{\ket{0}, \ket{\pm1}, \ket{\pm2}, \ket{\pm3}\}$, we can find a nearly MES, $\mathcal{K}=6.91$ and $\mathcal{F_{\text{glob}}}=0.39$. The resulting maxima are again outside of the target subspace and reduce the global fidelity.}
    \label{figureAppendix03neuW}
\end{figure*} 

For a pump $\text{LG}(1,0)$ the maximum amplitudes are $C^{0,0}_{0,1}=C^{0,0}_{1,0}$. Interestingly, the solution leads mainly to the state $\frac{1}{2} \bigl( \ket{0,-1;0,1} + \ket{0,1;0,-1} + \ket{1,0;0,0} + \ket{0,0;1,0}  \bigl)$, using the abbreviation $ \ket{p_{\signal},\ell_{\signal}} \otimes \ket{p_{\idler},\ell_{\idler}} :=  \ket{p_{\signal},\ell_{\signal};p_{\idler},\ell_{\idler}}$. However, this state is not maximally entangled within the subspace $\text{span}\{ \ket{p,\ell}\}, p=0,1, \ell=-1,0,1$. See also Sec. \ref{Appendix: Include higher radial indices of signal and idler} for further details.

Similarly, a pump superposition with different radial indices, e.g., with $p_{\pump}=0,1$, leads to a system of equations of the form
\begin{widetext}
    \begin{equation*}
\begin{pmatrix}
a_{0,0} \; C^{0,0}_{0,0}(0,0) + a_{1,0} \; C^{0,0}_{0,0}(1,0) & a_{0,1} \; C^{0,1}_{0,0}(0,1) + a_{1,1} \; C^{0,1}_{0,0}(1,1) & a_{0,2} \; C^{0,2}_{0,0}(0,2) + a_{1,2} \; C_{0,0}^{0,2}(1,2)  \\
a_{0,1} \; C^{1,0}_{0,0}(0,1) + a_{1,1} \; C^{1,0}_{0,0}(1,1) & a_{0,2} \; C^{1,1}_{0,0}(0,2) + a_{1,2} \; C^{1,1}_{0,0}(1,2) & a_{0,3} \; C^{1,2}_{0,0}(0,3) + a_{1,3} \; C_{0,0}^{1,2}(1,3)   \\
a_{0,2} \; C^{2,0}_{0,0}(0,2) + a_{1,2} \; C^{2,0}_{0,0}(1,2) & a_{0,3} \; C^{2,1}_{0,0}(0,3) +  a_{1,3} \; C^{2,1}_{0,0}(1,3) & a_{0,4} \; C^{2,2}_{0,0}(0,4) + a_{1,4} \;C_{0,0}^{2,2}(1,4) \\
\end{pmatrix}
\overset{!}{=} \frac{1}{\sqrt{3}} \begin{pmatrix}
1 & 0 & 0 \\
0 & 0 & 1 \\
0 & 1 & 0 \\
\end{pmatrix}. \\
\end{equation*}
\end{widetext}
In this case, we can set again $a_{0,1}=a_{1,1}=a_{0,2}=a_{1,2}=a_{0,4}=a_{1,4}=0$, but there are infinitely many solutions for the remaining pump weights. For example, choosing $a_{0,0} = 0.95\cdot \bigl(\sqrt{3} \: C^{0,0}_{0,0}(0,0) \bigl)^{-1}$, $ a_{1,0} = 0.05 \cdot \bigl(\sqrt{3} \: C^{0,0}_{0,0}(1,0) \bigl)^{-1}$, $ a_{0,3} =   0.95\cdot \bigl(\sqrt{3} \: C^{2,1}_{0,0}(0,3) \bigl)^{-1}$, and $a_{1,3} = 0.05 \cdot \bigl(\sqrt{3} \: C^{2,1}_{0,0}(1,3) \bigl)^{-1}$ yields the result shown in Fig. \ref{figureAppendix03neuW} (a). For the same reason as before, we conclude that the solution $a_{1,0}=a_{1,3}=0$ is the most efficient choice.


This also clarifies the dimensionality limit of MES in continuous subspaces discussed in the main text. For the subspace $\mathcal{S}_7=\text{span}\{ \ket{-3},...,\ket{3} \}$, we showed in Fig. 2 (g) of the main text an OAM spectrum, where the center modes along the anti-diagonal $\ell_{\pump}=0$ need to be amplitude-equalized, $C^{-1,1}_{0,0}=C^{0,0}_{0,0}=C^{1,-1}_{0,0}$.  Fig.~\ref{figureAppendix03neuW} (b) shows the OAM spectrum of an MES $\frac{1}{\sqrt{7}} \bigl( \ket{-3,-2} + \ket{-2,-3} + \ket{-1,1} + \ket{0,0}  + \ket{1,-1} + \ket{2,3} + \ket{3,2}  \bigl)$ with Schmidt number $\mathcal{K}=6.91$ obtained for a pump superposition additionally including the terms $ a_{1,0} \: \text{LG}_{1,0}, \: a_{2,0} \: \text{LG}_{2,0}, \: a_{3,0} \: \text{LG}_{3,0}$ to equalize the three modes along the anti-diagonal $\ell_{\pump}=0$. When viewed globally, the fidelity is reduced since the global maxima are the modes with amplitudes $C_{1,0}^{0,0}=C_{0,1}^{0,0}$. Increasing the dimensionality $d$ further leads to increasingly complex scenarios and systems of equations. Inverse design algorithms, such as those shown in \cite{karan2023postselection, comer2026inverse}, provide a powerful alternative in such cases, but come at the cost of losing the analytical simplicity. Our approach, based on disassembled subspaces, is an efficient alternative. \\


When calculating pump weights for a desired superposition in the broadband regime, the spectral dependence of the coincidence amplitudes must be taken into account, since $C^{\ell_{\signal},\ell_{\idler}}_{p_{\signal},p_{\idler}} (p_{\pump},\ell_{\pump}, w_{\pump})$ depends on $\Omega_{\signal}$ and $\Omega_{\idler}$ \cite{bernecker2023spatial, sevilla2024spectral}. To generate a state analogous to that in Fig. \ref{figureAppendix02neuW}(a), with equalized modes across different anti-diagonals in the broadband regime, the spectral dependence of the coincidence amplitudes must be made explicit, leading to
\begin{align*}
    a_{0,0} &=  \bigl(3 \int d\Omega_{\signal} d\Omega_{\idler} \; |C^{0,0}_{0,0}(0,0,w_{\pump}, \Omega_{\signal},\Omega_{\idler})|^2 \bigl)^{-1/2}
\end{align*}
and
\begin{align*}
    a_{0,3} &=  \bigl( 3 \int d\Omega_{\signal} d\Omega_{\idler} \; |C^{1,2}_{0,0}(0,3, w_{\pump},\Omega_{\signal},\Omega_{\idler})|^2 \bigl)^{-1/2}
\end{align*}
where the spectral filter bandwidth restricts the integration over the measured frequency band. 

In the broadband regime, these weights equalize the frequency-integrated populations of the selected OAM modes. This condition alone does not automatically guarantee a pure spatial MES: after tracing over frequency, the coherence between target terms is determined by the overlap of their joint spectral amplitudes. Nevertheless, Ref. \cite{baghdasaryan2022generalized} showed that all mode pairs sharing the same relative mode number, $N_R = 2p_{\pump}- 2p_{\signal}- 2p_{\idler} + |\ell_{\pump}| - |\ell_{\signal}| - |\ell_{\idler}|$, possess identical spectral dependence, such that the spatial and spectral degrees of freedom are separable. Since all target modes considered and suggested in our work satisfy $N_R = 0$, the resulting maximally entangled states remain spatially pure. 

\vspace{1cm}

\section{Comparison with large pump waists}
\label{Appendix: Comparison with large pump waists}

\begin{figure*}
    \centering
    \includegraphics[width=0.70\linewidth]{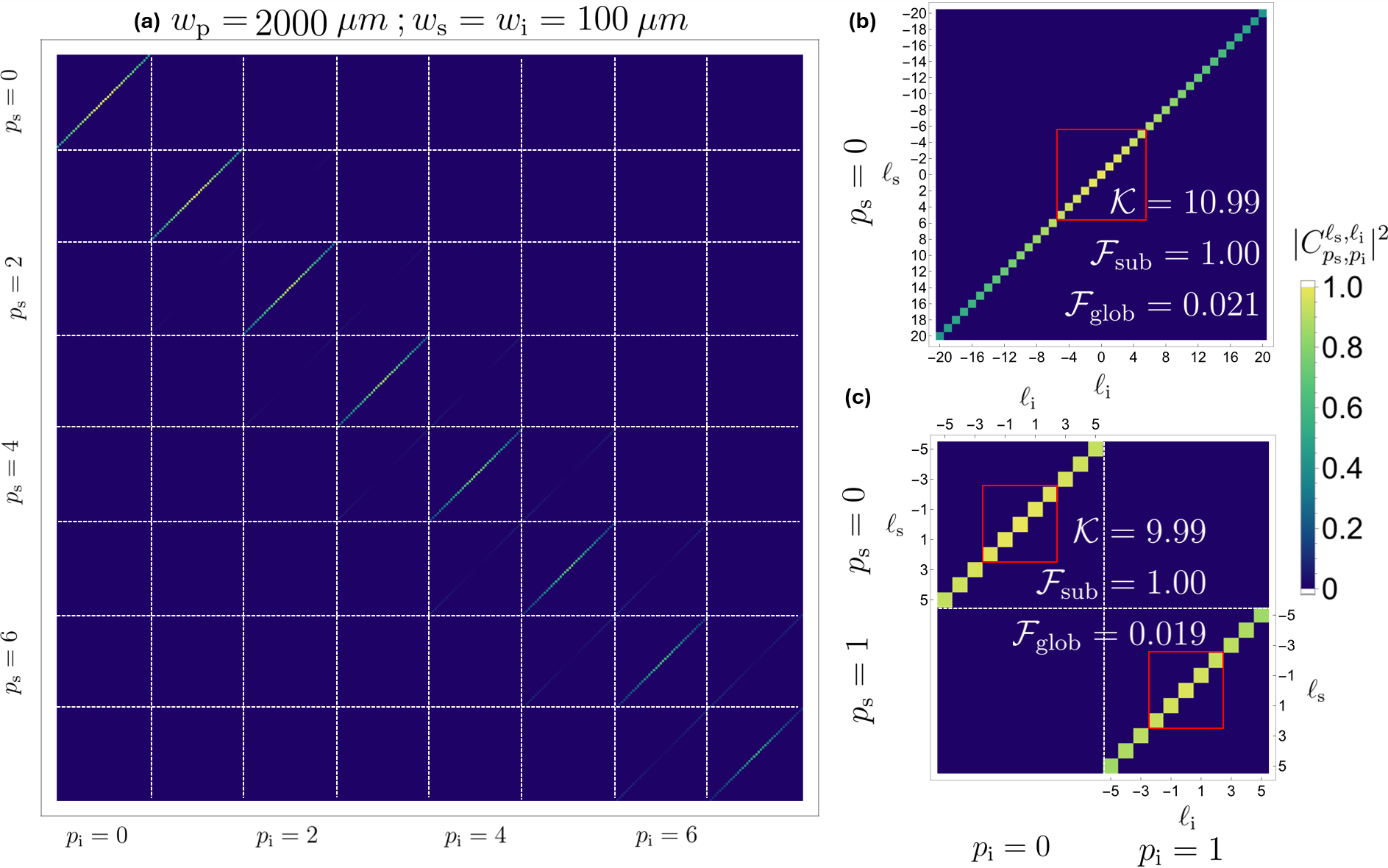}
    \caption{OAM spectra for large pump beam waist. (a) The pump beam carries zero OAM $\ell_{\pump}=0$. For a large waist ratio  $\gamma=w_{\pump}/w_{\signal}=w_{\pump}/w_{\idler}=20$, the spiral bandwidth is substantially broadened and more uniformly distributed. Non-negligible amplitudes are even maintained for high radial indices $p_{\signal}$ and $p_{\idler}$. (b) If we select the subspace $\mathcal{S}_{11}=\mathrm{span} \bigl\{ \ket{-5},\ket{-4}, ..., \ket{5} \bigl\}$ with $p_{\signal}=p_{\idler}=0$, we can obtain a $11 \times 11$ MES. However, globally, the large number of undesired modes outside the subspace renders this an inefficient method for preparing MESs. (c) We can alternatively include radial modes $p_{\signal}=p_{\idler}=1$ within the subspace of OAM modes $\text{span} \{ \ket{-2},..., \ket{2}\}$ since the spiral bandwidth is so broad, the central modes are matched too. Maximal entanglement within the selected subspace is achieved. Nevertheless, the global fidelity remains low due to the significant population outside the target subspace.}
    \label{figureAppendix05neuW}
\end{figure*}

If the pump beam waist $w_{\pump}$ is chosen much larger than the projection waists of the signal and idler $w_{\signal}=w_{\idler}$, the spiral bandwidth can be significantly broadened, leading to a more uniformly distributed OAM spectrum. In this section, we briefly compare our approach of generating MESs with broadening the OAM spectrum through a large pump waist.

The waist configuration is commonly characterized by the ratio  $\gamma=w_{\pump}/w_{\signal}=w_{\pump}/w_{\idler}$. For the SSMR, this ratio needs to be close to $\gamma=\frac{1}{\sqrt{2}}$. As an illustrative example, we consider a Gaussian pump ($\ell_{\pump}=0$) with a ratio $\gamma=20$, as shown in Fig. \ref{figureAppendix05neuW} (a). Compared with all previous examples, the resulting spiral bandwidth is substantially broader and remains significant even for higher radial indices $p_{\signal}=p_{\idler}$. As before, we can identify subspaces in which the modes are uniformly distributed. Since all modes are confined to a single anti-diagonal, the conditions for an MES are automatically satisfied if the amplitudes are equalized. We select two examples: one MES restricted to the zeroth radial mode and a second MES including higher radial indices.

In Fig. \ref{figureAppendix05neuW} (b), we consider the subspace  $ \mathcal{S}_{11}=\mathrm{span} \bigl\{\ket{-5},\ket{-4}, ..., \ket{4}, \ket{5} \bigl\}$ and the state
\begin{equation*}
    \ket{\psi} = \frac{1}{\sqrt{11}} \bigl( - \ket{-5,5} + \ket{-4,4} - ... + \ket{4,-4} -  \ket{5,-5}   \bigr).
\end{equation*}
Within the subspace, the modes are equally distributed, as quantified by a Schmidt number of $\mathcal{K}=10.99$ and a fidelity of $\mathcal{F_{\text{sub}}}=1.00$. Because the broadened spiral bandwidth also equalizes the amplitudes for modes with $p_{\signal}=p_{\idler}=1$, we can enlarge with respect to higher $p$ modes. 
The state
\begin{widetext}
    \begin{equation*}
    \ket{\psi} = \frac{1}{\sqrt{10}} \bigl( \ket{0,-2;0,2} - \ket{0,-1;0,1} + ... + \ket{0,2;0,-2} + \ket{1,-2;1,2} - \ket{1,-1;1,1} + ... +  \ket{1,2;1,-2} \bigr)
\end{equation*}
\end{widetext}
is maximally entangled in the subspace 
$\mathcal{S}_{10}=\mathrm{span} \bigl\{ \ket{0,-2},\ket{0,-1},\ket{0,0},  \ket{0,1}, \ket{0,2},  \ket{1,-2},\ket{1,-1}, \\ \ket{1,0},  \ket{1,1}, \ket{1,2} \bigl\}$
as evident from $\mathcal{K}=9.99$ and $\mathcal{F_{\text{sub}}}=1.00$. 

However, the global fidelities reveal that exploiting a broad spiral bandwidth is ultimately inefficient. While a larger pump waist increases the number of nearly equalized modes, it also distributes a substantial amount of the photon-pair modes into the long tails of the spectrum outside the chosen subspace. Consequently, the global fidelities are much smaller with only $\mathcal{F_{\text{glob}}}=0.011$ and $\mathcal{F_{\text{glob}}}=0.019$ for the considered two states in Fig. \ref{figureAppendix05neuW}, respectively. We note that in a broadband regime, the state is not spatially pure as the target modes have different RMNs. 

\section{Detailed description of Figure 4 (b)}
\label{Appendix: Detailed description of Figure 4 (b)}
Fig. 4 (b) in the main text summarizes the findings for MESs up to dimensionality $d=12$. Tab. \ref{tab:table2} lists the pump and subspace configurations selected using the design criteria of Sec. III B, together with their subspace fidelities, Schmidt numbers, and global fidelities. In particular, we specify which pump superpositions and subspace modes are recommended.

\begin{table*}
\caption{\label{tab:table2}
Data corresponding to Fig. 4 (b).}
\begin{ruledtabular}
\begin{tabular}{cccccccc}
$d=$ &$ V_{\pump} = \sum a_{\ell_{\pump}} \text{LG}_{0,\ell_{\pump}}, \ell_{\pump}=$ & $\mathcal{S}_d=\text{span}\{\ket{\ell}\},\ell=$ & $\mathcal{F}_{\text{sub}}$ & $\mathcal{K}$ & $\mathcal{F}_{\text{glob}}$ \\
\hline
3   & $0,3 $ & $0,1,2$ & $1.00$  & 3.00 & 0.679 \\
4   & $ \pm3$ & $\pm1,\pm2$ & 1.00  & 4.00 & 0.623  \\
5   & $0,\pm5$ & $0, \pm2,\pm3$ & 0.998 & 4.98 & 0.560 \\
6   & $\pm3,7$ & $\pm1,\pm2,3,4$ & 0.996 & 5.96 & 0.556 \\
7   & $\pm3,7,-6$ & $\pm1,\pm2,3,4,-3$ & 0.997 & 6.96 & 0.478 \\
8   & $\pm3,\pm7$ & $\pm1,\pm2,\pm3,\pm4$ & 0.995 & 7.91 & 0.525 \\
9   & $\pm3,\pm7,14$ & $\pm1,\pm2,\pm3,\pm4,7$ & 0.995 & 8.91 & 0.452 \\
10  & $\pm3,\pm7,15$ & $\pm1,\pm2,\pm3,\pm4,7,8$ & 0.994 & 9.90 & 0.473 \\
11  & $\pm3,\pm7,-14,15$ & $\pm1,\pm2,\pm3,\pm4,\pm7,8$ & 0.995 & 10.89 & 0.411 \\
12  & $\pm3,\pm7,\pm15$ & $\pm1,\pm2,\pm3,\pm4,\pm7,\pm8$ & 0.995 & 11.89 & 0.370 \\
\end{tabular}
\end{ruledtabular}
\label{GlobFidTab}
\end{table*}

\section{Include higher radial indices of signal and idler}
\label{Appendix: Include higher radial indices of signal and idler}

Recent advances in optical mode detection techniques, such as LG mode sorters \cite{fontaine2019laguerre}, have enabled the resolution of both $p$ and $\ell$ mode indices, for full modal characterization of LG states. Experimental demonstrations have reported efficient sorting for radial 
and OAM modes using various strategies, such as those based on the Gouy phase \cite{gu2018gouy}, or fractional Fourier transforms \cite{zhou2017sorting, fu2018realization}. Therefore, it is also of interest to explore the possibility of generating MES in subspaces that include both $\ell$ and higher $p$ modes.

The established design approach based on disassembled subspaces can be extended to subspaces that cover higher radial modes. Instead of enlarging the disassembled subspace through successively higher pump OAM values, one may alternatively introduce higher radial orders into the LG superposition. However, since no conservation law exists for $p$ modes, the generated distribution of radial signal and idler modes is less intuitive. Consequently, selecting suitable modes to form the subspace becomes less obvious since the dominant amplitudes cannot be identified a priori. Nevertheless, we observe a systematic trend that provides a useful guideline: in the SSMR for even $p_{\pump}$, the dominant amplitude occurs in the mode sector with $p_{\signal}=p_{\idler}=\frac{p_{\pump}}{2}$.  We now again use the notation $\ket{{p_{j},\ell_{j}}}$ for a single photon $j=\signal, \idler$ and $  \ket{p_{\signal},\ell_{\signal};p_{\idler},\ell_{\idler}}$ for two photons.

\begin{figure}[t]
    \centering
    \includegraphics[width=0.9\linewidth]{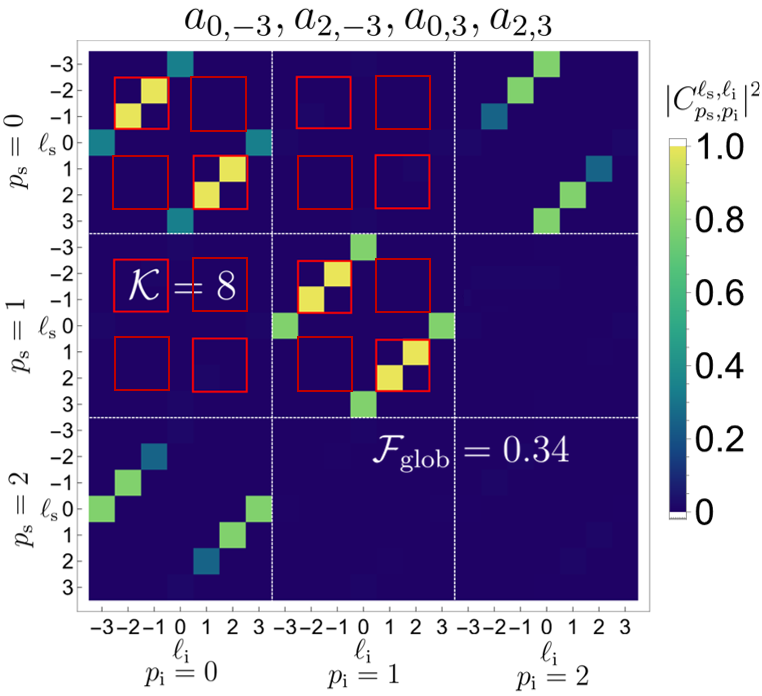}
    \caption{OAM spectra of a $d=8$ MES in a subspace with radial indices for signal and idler $p_{\signal}=p_{\idler}=0,1$. A pump superposition containing both OAM and radial modes enables a subspace-optimized MES whose target modes extend into sectors with nonzero signal and idler radial indices.}
    \label{figureAppendix04neuW}
\end{figure}

We look at an example for $d=8$ and start with the same OAM indices as in Fig. 4 of the main text for the single-photon subspace $\mathcal{S}_4=\text{span}\{ \ket{0,\pm1},\ket{0,\pm2} \}$. The modes generated by OAM pump modes $a_{0,\pm3}$ are still located in the upper-left sector of the OAM spectrum. Allowing $p_{\signal}=p_{\idler}=0,1$ effectively doubles the dimensionality. The total subspace $\mathcal{S}_8=\text{span}\{ \ket{0,\pm2},\ket{0,\pm1},\ket{1,\pm1},\ket{1,\pm 2} \}$ is highlighted in Fig. \ref{figureAppendix04neuW} in red. By adding suitable weights $a_{2,\pm3}$, we generate the most probable modes $C^{1,2}_{1,1}=C^{2,1}_{1,1}$ and $C^{-1,-2}_{1,1}=C^{-2,-1}_{1,1}$ in the subspace part with $p_{\signal}=p_{\idler}=1$. 

We have to check that undesired, non-vanishing amplitudes within the sector $p_{\signal}=0, p_{\idler}=1$ and $p_{\signal}=1,p_{\idler}=0$ do not fall within the constructed two-photon Hilbert space, as this would reduce the Schmidt number and subspace fidelity. Employing radial pump modes with $p_{\pump} \neq 0$ increases the number of non-negligible contributions of signal and idler modes with higher radial indices. Compared to the states shown in Fig. 4 (b) of the main text, this reduces the global fidelity. This trend becomes more pronounced as $d$ is further increased and, therefore, less attractive for an efficient experimental preparation of MESs.

\bibliography{apssamp}

\end{document}